# The Reconstruction Algorithm Study of 2D Interpolating Resistive Readout Structure

XIU Qing-Lei(修青磊)[1, 2, 3] DONG Ming-Yi(董明义)[1,2] LIU Rong-Guang(刘荣光)[1, 2] ZHANG Jian(张建)[1, 2] OU-YANG Qun(欧阳群)[1, 2] CHEN Yuan-Bo(陈元柏)[1, 2]

[1] State Key Laboratory of Particle Detection and Electronics, Beijing 100049, China
[2] Institute of High Energy Physics, CAS, Beijing 100049, China
[3] University of Chinese Academy of Sciences, Beijing 100049, China

**Abstract:** Systematic investigations including both simulation and prototype tests have been done about the interpolating resistive readout structure with GEM (Gaseous Electron Multiplier) detector. From the simulation, we have a good knowledge of the process of charges diffusion on the surface of the readout plane and develop several reconstruction methods to determine the hit position. The total signal duration time of a typical event with the readout structure is about several hundred nanoseconds, which implied an ideal count rate up to $10^6$Hz. A stable worked prototype was designed and fabricated after the simulation. Using $^{55}$Fe 5.9keV X-Ray, the image performance of the prototype is examined with flat field image and some special geometry shapes, meanwhile, an energy resolution of about 17% is obtained.

## 1. Introduction

For Micro Pattern Gaseous Detectors, such as GEM (Gaseous Electron Multiplier)[1] and Micromegas[2], pixel pad is the most suitable readout structure. However, in order to obtain a good space resolution, one has to reduce the size of the pixel and employ a large amount of pixels to cover a demanded effective area, which leads to a dramatically increasing of electronic channels. The virtual-pixel detector implemented with the two-dimensional interpolating resistive readout structure [3][4][5] which was developed by H. J. Besch can provide spatial resolution of about several hundred micro meters with an enormous reduction of electronic channels compared to pure pixel detectors. This readout structure that has high space resolution and low number of electronic channels is always a good option for many real applications.

With this concept, we do some numerical simulations and developed a GEM detector with two-dimensional interpolating resistive readout structure to have a good understand on its performances.

## 2. Basic principles

### 2.1 Diffusion and collection of charge

We combined the resistive readout structure with a triple-GEM gas gain device in our detector development [6]. The electron cluster, generated by ionizing radiation and multiplied in an avalanche process in the GEM detector, will drift to the surface of 2D interpolating readout structure in the electric drift field, and then the charges will diffuse on the resistive plane and be collected by the readout nodes of the adjacent cells. Using the information of the collected charges, the hit position can be reconstructed with suitable algorithm.

With thick film process, the readout plane is assembled by cells consisting of a high resistivity square pad and low resistivity narrow strips, which can obtain fast time respond and good position resolution. Each crossing point of the low resistive strips, so-called readout node, is connected to a low input impedance readout channel. Combined with the nearest metal foil coated on the lower surface of GEM film, the readout structure can be treated as a distributed 2-dimensional resistive-capacitive network. The schematic of the resistive anode is illustrated in Fig.1.

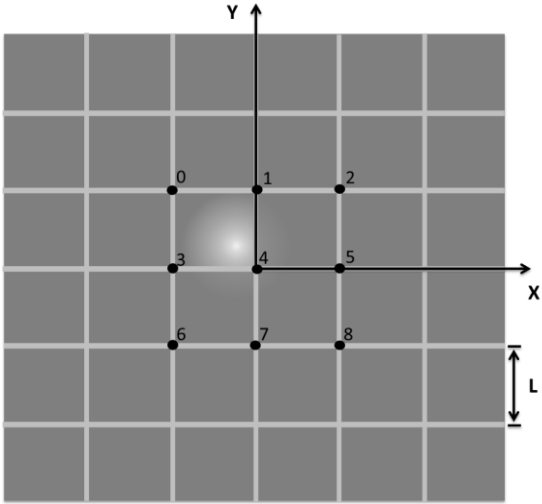

Fig. 1. Schematic illustration of resistive anodes. The edge length of a cell amounts to L=8mm. The surface resistance of the pad and strip is 100kΩ/□ and 1kΩ/□ respectively. Every crossing point of low resistive strip stands for a readout node. To illustrate the basic reconstruction algorithms, 9 readout nodes are represented by black point. The white point with halo stands for a typical event which is reconstructed with following algorithms.

### 2.2 Basic reconstruction method

The position of the incident particle has an approximate linear relationship with charges ($Q_i$) collected at the nodes of the adjacent cells. The simplest and most obvious linear reconstruction methods are so-called 4-, 6- and 3-node reconstruction algorithms. The node indications and the coordinate systems of the particular algorithms are also given in Fig1. The specified algorithms are listed below: [7]

4-Anode Algorithm: The origin of the coordinate system is located at anode 3.

$$x_4 = L \frac{(Q_1+Q_4)-(Q_0+Q_3)}{Q_{\tau 4}} \quad (1)$$

$$y_4 = L \frac{(Q_0+Q_1)-(Q_3+Q_4)}{Q_{\tau 4}} \quad (2)$$

$$Q_{\tau 4} = Q_0 + Q_1 + Q_3 + Q_4 \quad (3)$$

6-Anode Algorithm: Anode 4 is the origin.

$$x_6 = L \frac{(Q_2+Q_5)-(Q_0+Q_3)}{Q_{\tau 6x}} \quad (4)$$

$$y_6 = L \frac{(Q_0+Q_1)-(Q_6+Q_7)}{Q_{\tau 6y}} \quad (5)$$

$$Q_{\tau 6x} = Q_0 + Q_1 + Q_2 + Q_3 + Q_4 + Q_5 \quad (6)$$

$$Q_{\tau 6y} = Q_0 + Q_1 + Q_3 + Q_4 + Q_6 + Q_7 \quad (7)$$

3-Anode Algorithm: The origin is the same as 6-Anode Algorithm.

$$x_3 = L \frac{Q_5 - Q_3}{Q_{\tau 3x}} \quad (8)$$

$$y_3 = L \frac{Q_1 - Q_7}{Q_{\tau 3y}} \quad (9)$$

$$Q_{\tau 3x} = Q_3 + Q_4 + Q_5 \quad (10)$$
$$Q_{\tau 3y} = Q_1 + Q_4 + Q_7 \quad (11)$$

## 3. Simulation results

A numerical simulation was done to investigate the physical behaviors of 2D interpolating readout structure in our pre-research stage. With the simulation results, we can verify the reconstruction algorithms and know about the respond time of this structure.

### 3.1 Mathematic model

In one dimension the resistive anode can be treated as series connection of many identical small resistors which are parallel connected by many small capacitances formed by the metal foil of GEM and the surface of resistive anode. The cross connection of the resistive-capacitive chain forms a two-dimensional network. The schematic of the model is illustrated in Fig.2.

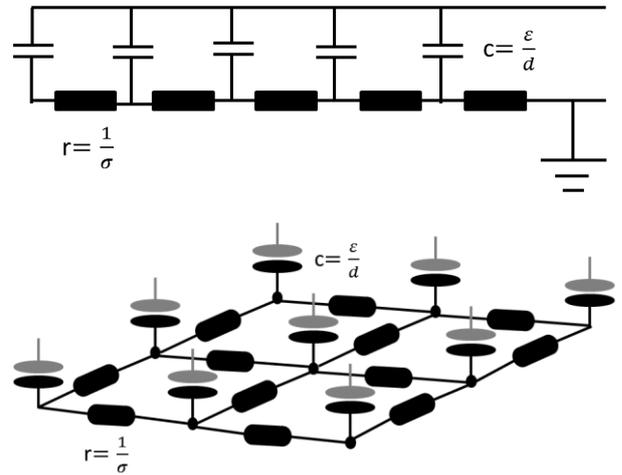

Fig. 2. Resistive-capacitive network model of the interpolating resistive readout structure. The continuous resistive plane is subdivided into small resistances which form a network. At every crossing node, a small capacitance contributed by GEM foil is connected. The simulation equation is derived from this model.

The equation describing the time evolution of the surface charge density function when the charge is deposited on the resistive anode can be derived from Ohm's law and the conservation of electronic charge in this model.

Because the resistive anode is made of two different resistive materials, the resistance of the anode is spatial dependent. In fact, besides the metal foil of the GEM film, there are some metal wires behind the resistive anode board. However, the capacitance induced by the backside wire is very small and can be neglected to simplify the simulation.

$$\vec{j} = \sigma(x,y) \cdot \vec{E} = -\sigma \cdot \nabla V \tag{12}$$

From the model in Fig.2, denoting the constant capacitance by c, one can obtain the surface charge density:

$$\rho(x,y,t) = c \cdot V(x,y,t) \tag{13}$$

Inserting Eq. (13) into equation (12), one can obtain:

$$\vec{j} = -\frac{\sigma}{c} \cdot \nabla \rho \tag{14}$$

$$\nabla \cdot \vec{j} = -\frac{\partial \rho}{\partial t} \tag{15}$$

Together with Ohm's law (14) and the conservation of electronic charge (15) the time derivative of the charge density can be obtained.

$$\frac{\partial \rho}{\partial t} - \frac{1}{c}[(\nabla \sigma) \cdot (\nabla \rho) + \sigma \nabla^2 \rho] = I(x,y,t) \tag{16}$$

I(x, y, t), which is assumed to be 2D Gaussian distribution in spatial part and 1D Gaussian distribution in time, is used to simulate the current signal caused by incoming particles.

Because the input impedance of the electronic channel is much less than the resistance of the resistive strips, the readout nodes can be treated as ideal drains approximately, where the charge density is equal to zero at all times. This is regarded as the boundary condition of equation (16).

The analytic solution of this equation can't be obtained with this boundary condition. We transform the partial differential equation to difference equation which can be solved with numerical method.

To obtain the difference equation, the coordinates and time are subdivided into equal intervals Δx, Δy and Δt.[8] The space-time coordinate of one grid point is (i*Δx, j*Δy, k*Δt). The schematic of the grid network is shown in Fig. 3. Every grid point is associated with particular properties such as $\rho_{i,j,k}$ and $\sigma_{i,j}$. In order to take into account the cross diffusion of charges to neighbor cells and compare different reconstruction methods, 3*3 cells are simulated. The time evolution of charge density can be transformed from equation (16):

$$\begin{aligned}\rho_{i,j,k+1} &= \rho_{i,j,k} \\&+ \frac{\Delta t}{2c\Delta x^2}\{[(\sigma_{i+1,j} + \sigma_{i,j})(\rho_{i+1,j,k} - \rho_{i,j,k}) \\&- (\sigma_{i,j} + \sigma_{i-1,j})(\rho_{i,j,k} - \rho_{i-1,j,k})] \\&+ [(\sigma_{i,j+1} + \sigma_{i,j})(\rho_{i,j+1,k} - \rho_{i,j,k}) \\&- (\sigma_{i,j} + \sigma_{i,j-1})(\rho_{i,j,k} - \rho_{i,j-1,k})]\} + I_{i,j,k}\Delta t\end{aligned}$$
(17)

$$\frac{\sigma_{max}\Delta t}{c\Delta x^2} \leq \frac{1}{4} \tag{18}$$

This equation can be resolved in numerical with a recursive algorithm. To have a stable resolution, formula (18) must be satisfied when determine the interval Δx and Δt [9].

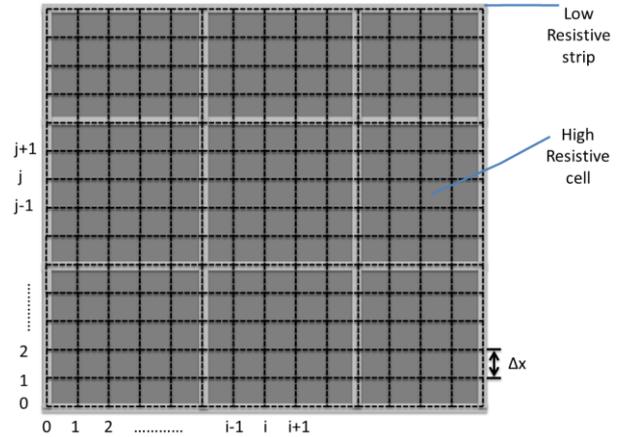

Fig. 3. Schematic of the grid network. The whole readout structure is subdivided into equal intervals by dashed lines. Every grid point is associated with particular properties such as $\rho_{i,j,k}$ and $\sigma_{i,j}$.

After the building of the mathematic model, we develop a program to solve the model. The simulation results are showed below.

### 3.2 Charge diffusion

Most of the charges are collected by the four nodes at the corner of the cell. However, it's obviously that some charges will diffuse to neighbor cells through the low resistive strips, which will lead to some distortion of the reconstructed position. According to the theory analysis, increasing the ratio of the pad resistivity to the strip resistivity can reduce the cross diffusion. If the ratio of the high resistivity to the low resistivity is infinity, all charges will be collected in one cell. Nevertheless, the high resistivity is limited to less than 1MΩ/□ by count rate capability and the low resistivity is restricted to more than 1kΩ/□ by the resistive noise in real condition. Fig.4 shows the time evolution of charge density on the surface of the resistive anode.

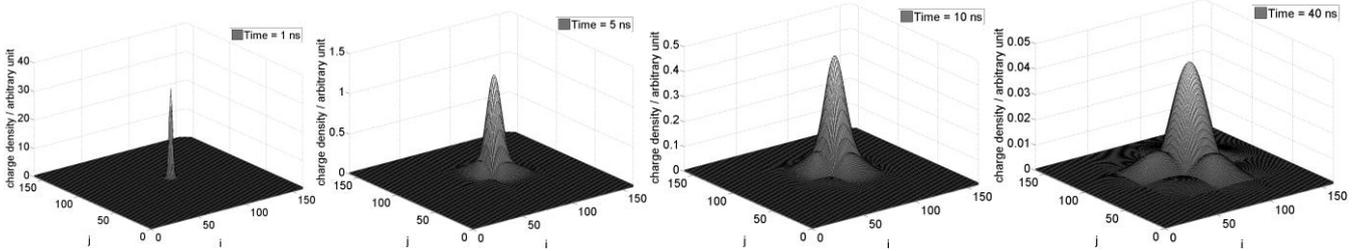

Fig. 4. Time evolution of the charge density. At the beginning, charges are concentrated in a very small region with a high charge density. With the diffusion and collection of the charges, the distribution of the charges becomes wider and more tenuous. From the last distribution graph, the low resistive strips and the readout nodes can be recognized clearly.

### 3.3 Time respond

The charges diffuse on the surface of the resistive plane with a time constant which is related to the resistivity of the plane and the capacitance between the GEM film and the resistive plane. The capacitance is fixed to be about 5nF/m$^2$ once the detector was manufactured with a distance of 2mm between the two layers. For the pad with resistivity of 100kΩ/□ and size of 8×8mm$^2$, the charge collection time is about 100ns. Considering the drift time of electron in the gas, the total time of one event in the detector is about several hundred nanoseconds [10]. Equipped with suitable electronics, the count rate of the system can approach up to 10$^6$Hz.

The current signals of events at different distances from the collecting node are shown in Fig.5; one can see that the duration time of the signal is about 100ns.

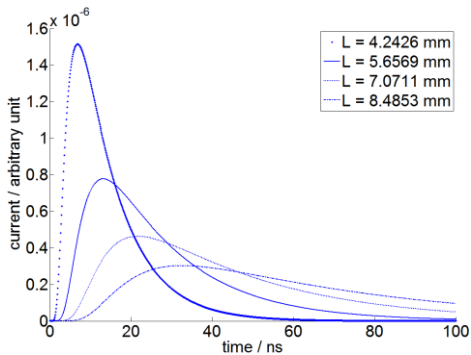

Fig. 5. Current signals of events at different distances from the readout node. The length of one cell is 8mm. According to this result, the majority of charges are collected in 100ns for high resistivity of 100kΩ/□.

### 3.4 Reconstruction

In the simulation, we can adjust the input position of the incoming current signal to form a regular pattern (fig.6(a)). With the reconstruction method mentioned before, we obtain the reconstructed pattern which can be compared with the regular input one.

In Fig.6 (b) ~ (d), the reconstructed positions of the 4-, 6- and 3-node algorithm are shown. The four-anode algorithm shows perfect reconstructions at the center of the cell, whereas the hits close to the strip are distorted. Contrast to the four-anode algorithm, both the three- and the six-anode algorithm have large distortions in the cell center, but they are more suitable algorithms for the hits close to borders of the cell because their distortions are much smaller than the four-anode algorithm. The similarity of reconstruction behavior of the 3- and the 6-node algorithm comes from the using of comparable symmetry planes and readout nodes.

In order to reconstruct the hit positions with fewer distortions, a new algorithm so-called weighted average algorithm is introduced (fig.6(e)). The following equations define this algorithm.

$$x_m = a_x x_4 + (1 - a_x)[b_x x_3 + (1 - b_x)x_6] \qquad (19)$$
$$y_m = a_y y_4 + (1 - a_y)[b_y y_3 + (1 - b_y)y_6] \qquad (20)$$

Since the position reconstructed from 3- and 6-node algorithm is very similar, we set $b_x=b_y=0.5$, which means that the weight of the 3- and the 6-node algorithm is equal. However, it's obviously that the weights $a_x$ and $a_y$ are spatial dependent. The weight of 4-node algorithm should be large at the center of the cell and become smaller and smaller approaching the borders. We find a piecewise continuous linear function to describe the spatial dependence of $a_x$, $a_y$. The graph of the function is shown in Fig.7. In real condition, the weights $a_x$ and $a_y$ should be tuned with a standard measurement firstly.

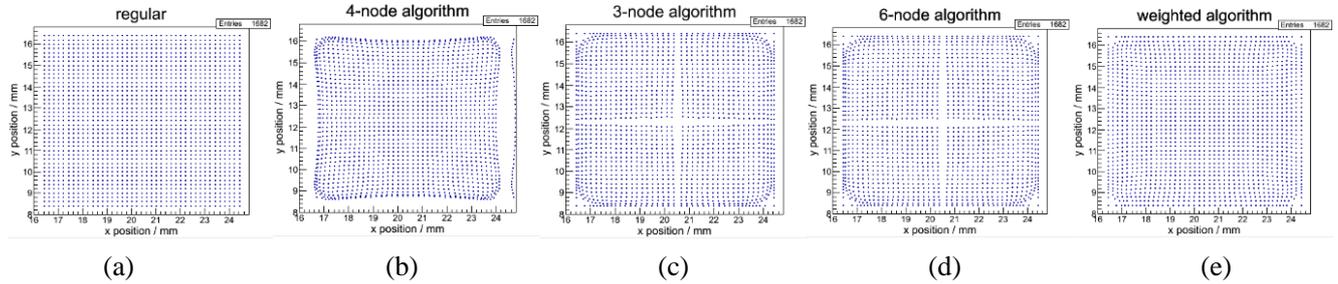

(a)　　　　　　(b)　　　　　　(c)　　　　　　(d)　　　　　　(e)

Fig. 6. Reconstruction positions with the simulation data in one pad. There are always distortions for simple reconstruction algorithms. To reduce the distortions, a weighted average algorithm is introduced.

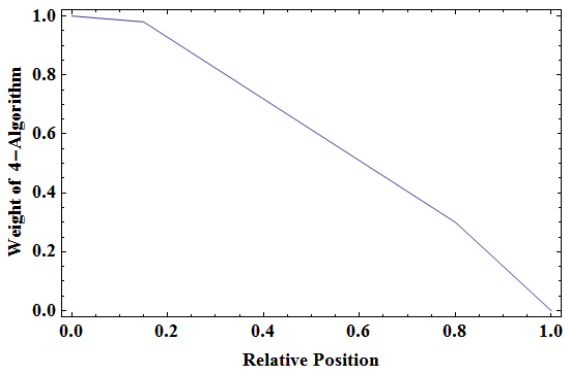

Fig. 7. The weight of 4-node algorithm in the weighted average algorithm. The x coordinate is the relative position of the event. 0 means the center of the cell and 1 means the border. The y coordinate stands for factors $a_x$ and $a_y$, the weight of 4-node algorithm, in equation (19) and (20). This function can be tuned with measure results.

## 4. Test results

The prototype with the resistive readout structure has been developed and has been tested with $^{55}$Fe X rays. The resistivity of the pad is set to be 100kΩ/□ and the strip to be 1kΩ/□, which has a fast respond time and keeps the charge losses to be small.

### 4.1 Reconstruction software

To process the large amount of data effectively, we developed analysis software with object-oriented technology in C++. The schematic of the software is illustrated in Fig.8. The software consist of these processes: find seeds, reconstruct positions and energy, save reconstructed data in ROOT files.

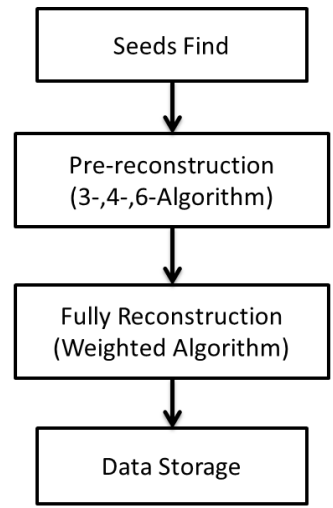

Fig. 8. Flow chart of data processing software. An iterative technique is used in the reconstruction

The basic topologic structure of seed consists of five cells and the charge is deposited on the central cell. Using the data in the seed one can obtain the position and energy information with different algorithms. For seeds located at the edge or the corner of the readout structure, the seed consists of four or three cells respectively.

Fig.9 shows the illustration of basic seed structure

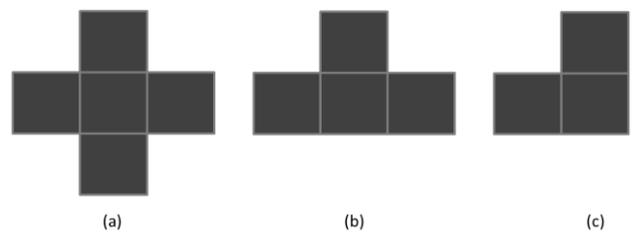

Fig. 9. Basic configurations of reconstruction seeds. Almost all information of one event can be obtained from the seed.

### 4.2 Image performance

The image performances of the prototype are studied using 3×3 cells with $^{55}$Fe 5.9keV X-ray. Firstly, a flat field image is obtained, which can be used to tuning the weighted-average algorithm. The piecewise function is adjusted until the best flat field image is obtained. After this step is finished, the algorithm can be stably used unless some parameters of the detector are changed.

The flat field image is shown in Fig.10. The image is reconstructed with all algorithms mentioned before. The test results are very similar to the simulation results. The lack of the events at the four corners is due to the trigger condition of the test system.

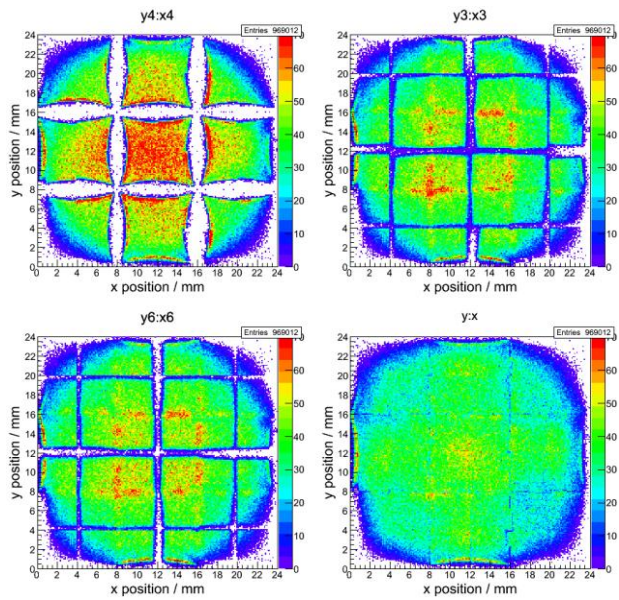

Fig. 10. Flat field images which is used to tuning the reconstruction software. The measured results and the simulation results are in good accordance, which indicate the correctness of the simulations. The absence of events at the four corners is caused by the trigger setup.

An image of a round hole reconstructed with the weighted-average algorithm is shown in Fig.11. All the parameters are obtained from the flat field tuning. The distortions close to the cell borders are suppressed to be very small.

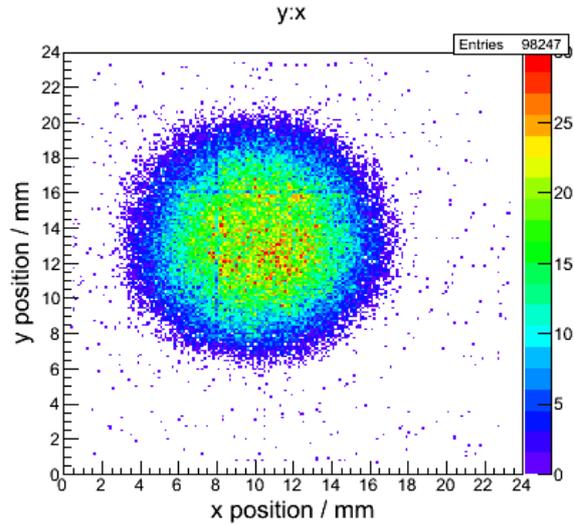

Fig. 11. The image of a small hole. Using the weighted average algorithm, the distortion closing to borders is hardly to be recognized.

### 4.3 Energy Resolution

Since the charges of one event can be fully collected, we can obtain the energy spectrum deposited in the detector.

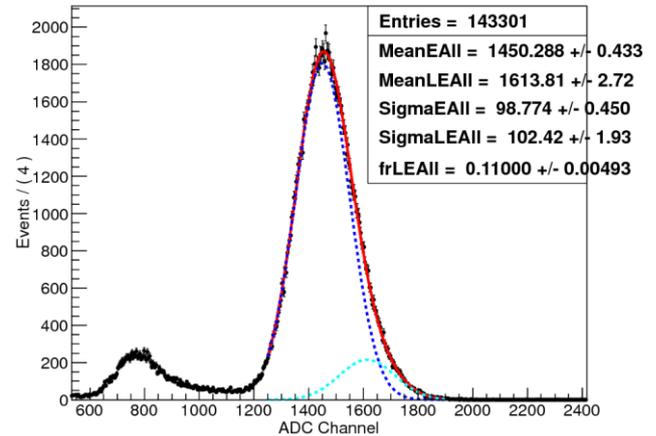

Fig. 12. Energy spectrum of $^{55}$Fe 5.9keV X-Ray. The energy resolution of the prototype is about 17.4%@5.9keV taking account of the influence of 6.49keV X-Ray which accounts for about 10.5% of the total photon flux.

The energy spectrum isn't an ideal Gaussian shape, because the X-Ray photons emitted from $^{55}$Fe consist of 5.9keV and 6.49keV level. The two levels are too close to separate in our detector. We use a double Gaussian function to fit the data. The energy resolution is about 17.4% at 5.9keV, and the ratio of 6.49keV peak to 5.9keV peak is 1.113±0.002, which is consistent to the theoretic value.

## 5. Conclusion

Systematic investigations about the 2-Dimensional Interpolating resistive readout structure have been done in the development of a GEM prototype.

A mathematic model was built to simulate the charge diffusion process at the resistive surface, and to determine the key parameters such as resistivity, pad scale of the resistive readout structure. For the pad with resistivity of 100kΩ/□ and size of 8×8mm$^2$, the total signal duration time of a typical event is about several hundred nanoseconds. This simulation results imply that the readout structure has potential to approach a count rate of $10^6$Hz.

Because the simple linear reconstruction algorithms that just use the charges collected by 4 nodes of a cell has defect for the hits close to border of the cell, a weighted average algorithm combining 3 basic linear algorithms is introduced. In this algorithm, the position is reconstructed with the 3, 4, 6-nodes method respectively at the first step and then an average position is obtained with different weights for 3 basic linear algorithms. With correct calibrations and optimizations, the distortion caused by charge loss close to the cell border can be suppressed to be very small.

The GEM prototype was tested with $^{55}$Fe X-Rays. Using the weighted average reconstruction algorithm, the prototype has a very good image performance, and the distortions of the images are very small, which is consistent to the simulation results. The test result also shows that the energy resolution of the prototype is about 17% when take into account the two energy levels of $^{55}$Fe X-rays.

From the simulations and prototype tests, we can see that the interpolating resistive readout structure has good time resolution and image performance with low number of electronic channels. With these advantages, widely applications of this interpolating readout structure for micro-pattern gaseous detector can be expected.

## Acknowledgements

*We are especially indebted to Prof. Xie Yi-gang, Dr. Fan-RuiRui and Dr. Lv-Xinyu for their useful suggestions and discussions.*